\documentclass[twocolumn, switch]{article} % Method A for two-column formatting

\usepackage{preprint}

\usepackage{amsmath, amsthm, amssymb, amsfonts}

\usepackage[numbers,square]{natbib}
\usepackage[utf8]{inputenc}	% allow utf-8 input
\usepackage[T1]{fontenc}	% use 8-bit T1 fonts
\usepackage{xcolor}		% colors for hyperlinks
\usepackage[colorlinks = true,
            linkcolor = purple,
            urlcolor  = blue,
            citecolor = cyan,
            anchorcolor = black]{hyperref}	% Color links to references, figures, etc.
\usepackage{booktabs} 		% professional-quality tables
\usepackage{nicefrac}		% compact symbols for 1/2, etc.
\usepackage{microtype}		% microtypography
\usepackage{lineno}		% Line numbers
\usepackage{float}			% Allows for figures within multicol

\usepackage{lipsum}		%  Filler text

\usepackage{newfloat}
\DeclareFloatingEnvironment[name={Supplementary Figure}]{suppfigure}
\usepackage{sidecap}
\sidecaptionvpos{figure}{c}

\usepackage{xcolor}
\usepackage{braket}
\usepackage[normalem]{ulem}
\usepackage{subcaption}
\usepackage{amsmath}

\usepackage{multirow}
\usepackage{rotating}
\usepackage{booktabs}
\usepackage[T1]{fontenc}
\usepackage{array} 

\usepackage{adjustbox}
\usepackage{romannum}
\usepackage{gensymb}
\usepackage{textcomp}
\usepackage[utf8]{inputenc}
\usepackage{titlesec}
\titlespacing\section{0pt}{12pt plus 3pt minus 3pt}{1pt plus 1pt minus 1pt}
\titlespacing\subsection{0pt}{10pt plus 3pt minus 3pt}{1pt plus 1pt minus 1pt}
\titlespacing\subsubsection{0pt}{8pt plus 3pt minus 3pt}{1pt plus 1pt minus 1pt}

\usepackage{tikz,xcolor,hyperref}

\definecolor{lime}{HTML}{A6CE39}
\DeclareRobustCommand{\orcidicon}{
	\begin{tikzpicture}
	\draw[lime, fill=lime] (0,0) 
	circle [radius=0.16] 
	node[white] {{\fontfamily{qag}\selectfont \tiny ID}};
	\draw[white, fill=white] (-0.0625,0.095) 
	circle [radius=0.007];
	\end{tikzpicture}
	\hspace{-2mm}
}
\foreach \x in {A, ..., Z}{\expandafter\xdef\csname orcid\x\endcsname{\noexpand\href{https://orcid.org/\csname orcidauthor\x\endcsname}
			{\noexpand\orcidicon}}
}
\title{Optimizing $\alpha''$-Fe$_{16}$N$_2$ as permanent magnet via alloying}

\usepackage{authblk}

\author[1]{Bo Zhao}
\author[1]{Ruiwen Xie}
\author[2\thanks{\tt{imants.dirba@tu-darmstadt.de}}]{Imants Dirba}
\author[3]{Lambert Alff}
\author[2]{Oliver Gutfleisch}
\author[1]{Hongbin Zhang}

\affil[1]{Theory of Magnetic Materials, Institute of Materials Science, Technical University of
Darmstadt, 64287 Darmstadt, Germany}
\affil[2]{Functional Materials, Institute of Materials Science, Technical University of Darmstadt,
64287 Darmstadt, Germany}
\affil[3]{Advanced Thin Film Technology, Institute of Materials Science, Technical University of Darmstadt,
64287 Darmstadt, Germany}

\begin{document}

\twocolumn[ % Method A for two-column formatting
  \begin{@twocolumnfalse} % Method A for two-column formatting
  
\maketitle

\begin{abstract}
Based on systematic first-principles calculations, we investigate the effects of 27 alloying elements on the intrinsic magnetic properties of Fe$_{16}$N$_2$, in order to further optimize its properties for permanent magnet applications. Analysis on the thermodynamic stabilities based on formation energy and distance to the convex hull reveals that 20 elements can be substituted into Fe$_{16}$N$_2$, where there is no strong site-preference upon doping. It is observed that all alloying elements can essentially reduce the saturation magnetization, whereas the magnetic anisotropy can be significantly modified. In terms of the Boltzmann- average intrinsic properties, we identify 8 elements as interesting candidates, with Co, Mo, and W as the most promising cases for further experimental validations.
\end{abstract}
%\keywords{First keyword \and Second keyword \and More} % (optional)
\vspace{0.35cm}

  \end{@twocolumnfalse}] % Method A for two-column formatting
 % Method A for two-column formatting

%\begin{multicols}{2} % Method B for two-column formatting (doesn't play well with line numbers), comment out if using method A

%%%%%%%%%%%%%%%  Main text   %%%%%%%%%%%%%%%
% \linenumbers

\section{Introduction}
The quest for high‐performance permanent magnets (PMs) has long been driven by the need for energy applications, ideally combining large coercivity, high energy product, and thermal stability without relying on critical or expensive elements~\cite{gutfleisch2011magnetic}. Standard rare‐earth–based high-performance PMs (e.g., $\mathrm{NdFeB}$- and $\mathrm{SmCo}$-based compounds)~\cite{sagawa1984new,holc1990study,becker1970rare,guo2022recent} can face supply‐chain constraints and cost volatility. 
%\textcolor{blue}{At least one reference per sentence}
As a result, there is intense interest in exploring rare‐earth–free alternatives, if not with comparable performance as the NdFeB/SmCo-based systems, as gap magnets~\cite{coey2012permanent,ochirkhuyag2024fe}. From a material design standpoint, intrinsic magnetic properties, i.e., saturation magnetization ($M_s$), magnetic‐anisotropy energy (MAE), and Curie temperature ($T_C$), set the upper limit of the extrinsic properties. For instance, the theoretical ceiling for $(BH)_{\mathrm{max}}$ is set by $J_S^2 / (4\mu_0)$, which limits to $\sim 1190\ \mathrm{kJ/m^3}$ even for anisotropic Fe-based compounds like $\mathrm{Fe_{0.65}Co_{0.35}}$~\cite{akai2018maximum,coey2010magnetism}.
To screen for proper PM candidates, high-throughput (HTP) density functional theory (DFT) calculations have been extensively applied~\cite{vishina2020high, zhang2021high, marathe2023exploration}.
Moreover, 
% \textcolor{blue}{Where is ``on the one hand"?}
substitutional alloying with judicious dopants can efficiently tailor MAE, $M_s$, and $T_C$ simultaneously, thus has been successfully applied to optimize conventional and rare-earth-free PMs. For instance, \citeauthor{gkouzia2023element} demonstrated that doping copper into SmCo$_5$ leads to an enhancement of anisotropy and a reduction of magnetization~\cite{gkouzia2023element}. Another illustrative case is the $\mathrm{Fe_2P}$ system, where Co and Si co-substitution raises $T_C$ above 500 K and achieves $K_1 \approx 1\ \mathrm{MJ/m^3}$, yielding a projected $(BH)_{\mathrm{max}}$ of 200 $\mathrm{kJ/m^3}$~\cite{he2022intrinsic}. 
% \citeauthor{golden2018evolution}~\cite{golden2018evolution} 
It is also noted that interstitial alloying with light elements can be applied to optimize the intrinsic magnetic properties, such as N-interstitials in Sm$_2$Fe$_{17}$~\cite{xia2019initial}, H-interstitials in SmCo$_5$~\cite{ye2020giant}, B-interstitials in Fe~\cite{golden2018evolution}, and light-element interstitials in Heusler compounds~\cite{gao2020designing}.
%influences tetragonalization in bcc Fe, distorted by interstitial boron, affecting its anisotropy.
%\textcolor{red}{Maybe here you can talk about the upper limit of permanent magnet as discussed in 10.1016/j.scriptamat.2018.02.006}
%\textcolor{blue}{Afterwards, it is a good idea to talk about strategies to design permanent magnets, in particular substitutional alloying as done in Fe2P https://advanced.onlinelibrary.wiley.com/doi/full/10.1002/adfm.202107513} 

% One of the potential candidates for permanent magnets, 
Iron‐nitride phases~\cite{coey1999magnetic,golden2017thin} have emerged as promising systems because they combine relatively low cost with tunable magnetic characteristics, within which one potential candidate is namely the $\alpha^{\prime\prime}$-$\mathrm{Fe_{16}N_2}$ phase~\cite{dirba2015increased,dirba2017synthesis}. The tetragonal distortion in $\mathrm{Fe_8N}$ creates a pronounced uniaxial MAE with $K_1=1.0 \mathrm{MJ/m^3}$, corresponding to an anisotropy field H$_a$=2K$_1$/J=1.1T (J$\approx$ 2.3T is the saturation polarization)~\cite{zhang2016engineering}. This makes $\mathrm{Fe_8N}$ inherently favorable for PMs
applications~\cite{wang2020environment, dirba2021synthesis,NironMagneticsDTIC,APNews2023}. However, its very high saturation magnetization ($\approx$ 1.75T)~\cite{zhang2016engineering} — while advantageous for maximizing magnetic flux — is actually a double‐edged sword: excessively large $M_s$ can lead to unwanted demagnetizing fields, complicate the domain‐wall pinning, and ultimately compromise the coercivity. In other words, although $\mathrm{Fe_8N}$ possesses the requisite anisotropy to maintain thermal stability of magnetization, its excessive $M_s$ drives the resulting magnet into a regime where self‐demagnetization become problematic.

% \textcolor{blue}{To mitigate this issue, likewise, one can introduce dilute concentrations of other transition‐metal dopants into the $\mathrm{Fe_8N}$ lattice. By substituting a fraction of the Fe sites with elements that carry a smaller magnetic moment, it becomes possible to dial back the total $M_s$ while preserving most of the intrinsic MAE. Moreover, certain dopants—especially those with strong spin–orbit coupling—can even enhance the uniaxial anisotropy further. The net effect is a more balanced set of magnetic parameters: a slightly reduced $M_s$ that avoids excessive demagnetizing effects, an MAE that remains large enough to sustain high coercivity, and a Curie temperature that stays well above room temperature. Strategic doping therefore offers a pathway to tailor $\mathrm{Fe_8N}$‐based compounds into genuine permanent‐magnet materials without sacrificing the advantages of a high‐iron content, low‐cost system.}
Introducing small amounts of transition-metal dopants into the Fe$_8$N lattice can slightly lower saturation magnetization (M$_s$) without significantly reducing magnetocrystalline anisotropy energy (MAE). Certain heavy-element dopants can even boost uniaxial anisotropy. The result is a material with reduced demagnetizing fields, high coercivity, and a Curie temperature well above room temperature—an iron-rich, cost-effective permanent magnet.
% \textcolor{red}{As I said, this should be shortened to one or two sentences!}
In this work, we employ DFT calculations to evaluate the intrinsic magnetic properties ($M_s$, MAE, and $T_C$) of transition‐metal–doped Fe$_{16}$N$_2$. By systematically substituting a selection of 3d, 4d, and 5d dopants for iron atoms
% \textcolor{red}{Plz revise after figuring out how ``subsitute" should be used properply. It is different from ``replace".}
to first assess thermodynamic stability and then explore the impact of the dopants on the delicate interplay between anisotropy and magnetization, we aim to identify compositions that maximize the energy product $(BH)_{\mathrm{max}}$
% \textcolor{red}{Plz be consistent with how you specified in the 1st paragraph.}
by achieving a high MAE while keeping $M_s$ at a level that ensures moderate demagnetizing field. The outcomes of this study provide guiding principles for the experimental realization of iron‐nitride–based permanent magnets that do not rely on rare‐earth elements.

\section{Computational Details}
Starting from the $\alpha^{\prime\prime}$-$\mathrm{Fe_{16}N_2}$ structure with the space group I4/mmm, we considered 6.25 at\%-doped 
Fe$_{15}$XN$_2$ with X being 27 transition metal elements (Sc, Ti, V, Cr, Mn, Co, Ni, Cu, Zn, Y, Zr, Nb, Mo, Ru, Rh, Pd, Ag, Cd, Hf, Ta, W, Re, Os, Ir, Pt, Au, Hg) except for Tc which is radioactive.
% \textcolor{blue}{3d/4d/5d = 10 x 3 = 30, - 1 = 29. Please specify the elements.}
The crystal structures of the alloyed systems were generated by replacing one of the iron atoms in $\alpha''$-$\mathrm{Fe_{16}N_2}$ by the corresponding X atoms. There are three symmetry nonequivalent iron sites (i.e., Wyckoff sites 4e, 4d, and 8h) in the Fe$_{16}$N$_2$ unit cell, resulting in $27\times 3 = 81$ different doped structures. Both lattice constants and atomic positions were optimized using the Vienna \textit{ab inito} Simulation Package (VASP)~\cite{kresse1996efficient,kresse1999ultrasoft} codes, with convergence tolerance for total energies and forces at $10^{-6}$ eV and $10^{-3}$ eV/\AA, respectively.  
The exchange-correlation functional was approximated using the Perdew-Burke-Ernzerhof parameterization~\cite{perdew1996generalized}, with a k-mesh of $8\times 8\times 7$ for structural optimization. Although the structures doped with 4e and 4d sites remained tetragonal (space group P4mm and P$\bar{4}$ m2, respectively), the 8h-doping results in an orthorhombic cell with the space group Amm2. However, we confirmed that the in-plane lattice constants remained the same in the optimized structures with the stress tensor $\sigma_{xx}\approx \sigma_{yy}$. The tetragonal geometry was therefore maintained for all the doping cases.
% \textcolor{blue}{Are the resulting space groups for 3-doped cases still tetragonal?}

As mentioned, there are essentially three intrinsic magnetic properties, i.e., $M_S$, MAE, and $T_C$,
for the identification of candidates as permanent magnets. In this work, MAE was evaluated based on the magnetic torque method~\cite{staunton2006temperature} using the SPRKKR~\cite{ebert2011calculating} code.
Under the magnetic torque theorem, the variation of magnetic free energy $F$, which is a function of magnetic direction determined by spherical angles $F(\phi, \theta)$, is equivalent to the calculation of torque given by
$\hat{T}(\hat{n}) = - \frac{\partial \hat{F}(\hat{n})}{\partial \hat{n}}$,
where $\hat{n} = (\phi, \theta)$ is the magnetization direction. As shown by Wang \textit{et al.}, for tetragonal cell with uniaxial magnets, $\hat{F}$ can be approximately parameterized as $\hat{F} \approx F_0 + K_1 \sin^2 \theta + K_2\sin^4\theta$, with $F_0$ being the isotropic term. Taking $\theta = 45\degree$, we end up with $\hat{T} = K_1 + K_2$, corresponding to the uniaxial MAE $= E_{[001]}-E_{[100]}$. $K_1$ in the following sections refers to MAE as a convention. Coherent potential approximation (CPA)~\cite{soven1967coherent} was used to simulate the configurational average for the alloying. Additionally, we conducted comparative VASP calculations to determine the MAE—computed as the energy difference between chosen magnetic orientations—and to break down the MAE into contributions from each atomic site.
In both cases, a dense k-mesh of $16\times16\times14$ was used to guarantee good convergence. Trends of MAE with respect to different dopants and different sites are consistent, despite a systematic discrepancy of absolute MAE values calculated between SPRKKR and VASP because of the different treatment of potential. The site-averaged SPRKKR calculations were used as the final results.
% \textcolor{blue}{Have you validated $M_S$ and MAE from SPRKKR and VASP codes?}

To evaluate $T_C$, Heisenberg exchange parameters $J_{ij}$ were calculated using SPRKKR based on the Lichtenstein formula \textit{et al.}~\cite{liechtenstein1984exchange}: 
$J_{ij}=
\frac{1}{4\pi}\,
\mathrm{Im} \int_{-\infty}^{E_F}
\mathrm{Tr}\bigl[\,
\Delta_i\,G_{ij}^{\uparrow}(E)\,
\Delta_j\,G_{ji}^{\downarrow}(E)
\bigr]
\,\mathrm{d}E,$
where $\Delta_i$ is the on-site exchange splitting operator and $G_{ij}^{\sigma}$ is the Green's function between sites i and j of spin channel $\sigma$. Two times the length of the unit cell was selected as the cluster radius for atomic pairs. The Curie temperature $T_C$ was then obtained by the mean-field approximation given by $T_C^{\mathrm{MFA}} = \frac{2}{3N}\sum_{i,j}J_{ij}$, where N is the number of the magnetic atoms in the unit cell.

\section{Results and Discussion}
\label{sec:results}

\begin{figure}[H]
  \centering
  \includegraphics[width=0.45\textwidth]{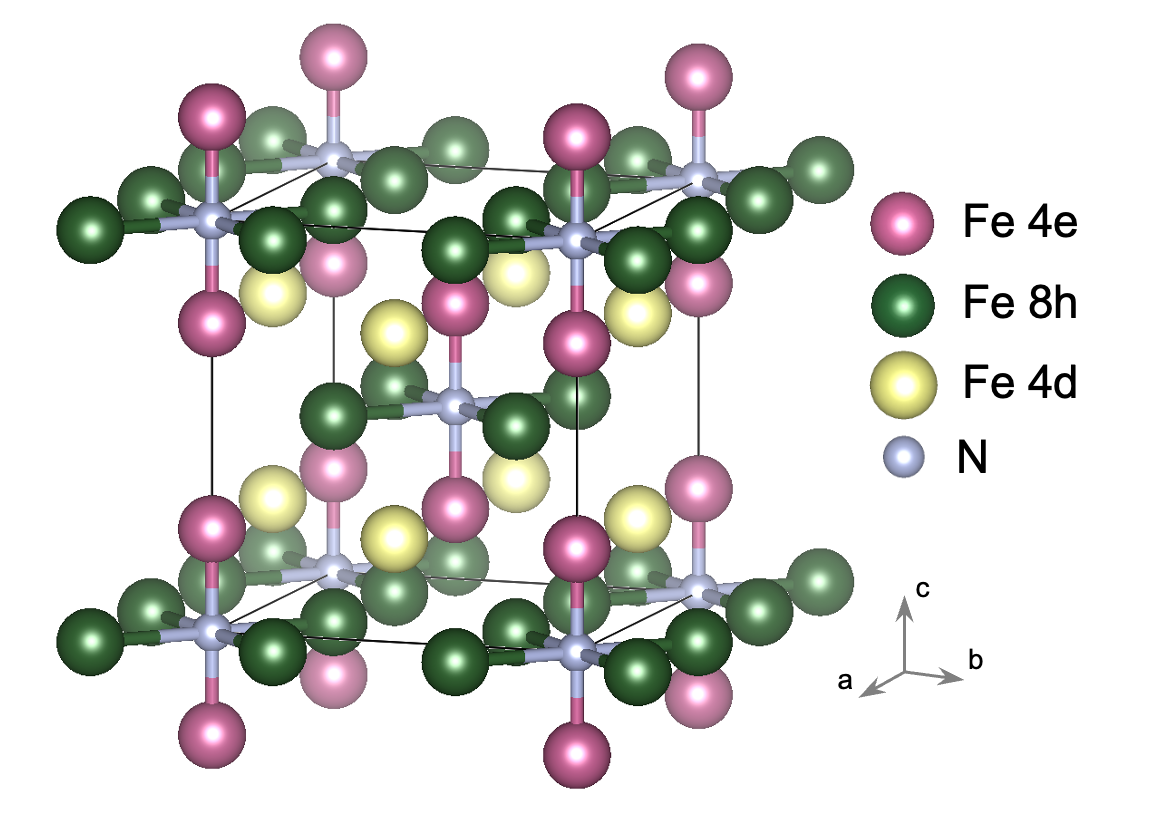}
  \caption{The crystal structure of $\alpha''\mathrm{Fe_{16}N_2}$ with space group I4/mmm (No. 139), with three different iron sites indicated by different colors.}
  \label{fig:struct}  
\end{figure}

As discussed previously~\cite{zhang2016engineering,golden2018evolution,gao2020designing}, the tetragonalization of the bcc crystal structure induced by interstitial N atoms at the octahedral centers plays an essential role for uniaxial MAE in Fe$_8$N. The site-resolved magnetic moments of pristine $\mathrm{Fe_{16}N_2}$ are 2.14, 2.82, and 2.35 $\mu_B$ for iron atoms on 4e, 4d, and 8h sites, respectively. Magnetic moments in 4e and 8h sites are comparable with that of $\alpha$-Fe, while a significant enhancement is observed for the 8h sites as a result of the reduction of occupancy in the minority spin channels (Figure S1). In terms of MAE $\propto \Delta \mu_L$ with $\Delta \mu_L$ being the change of orbital moments for two different magnetization directions~\cite{bruno1989tight}, 4e and 4d iron atoms favor the out-of-plane magnetization direction with $\Delta \mu_L$ being 20 and 4 m$\mu_B$, respectively; Whereas the iron atoms on the 8h sites favor the in-plane direction with $\Delta \mu_L$ = -4 m$\mu_B$. The density of states (DOS) along quantization direction [001] in Figure S1 (a)-(c) show that for the iron atoms on the 4e sites, the high $d_{xz\downarrow}$ and $d_{yz\downarrow}$ density just above $\mathrm{E_F}$ contribute to the non-vanishing matrix element of angular momentum quantized at z axis, i.e., $\langle xz| L_z| yz\rangle = 1$, giving rise to large orbital moment $\mu_L$ of 0.1 $\mu_B$. In contrast, for the iron atoms on the 8h sites, states around $\mathrm{E_F}$ are dominated by $d_{xy\uparrow}$, leading to finite value for angular momentum $L_x$. The energy shift between out-of-plane and in-plane magnetization direction can be more explicitly observed in the band structure shown in Figure S1 (d). Correspondingly, it is expected that doping on the 4d sites may result in the most efficient reduction of saturation magnetization and that on 8h sites leads to the most enhancement of MAE.

\begin{figure}
  \centering
  \includegraphics[width=0.48\textwidth]{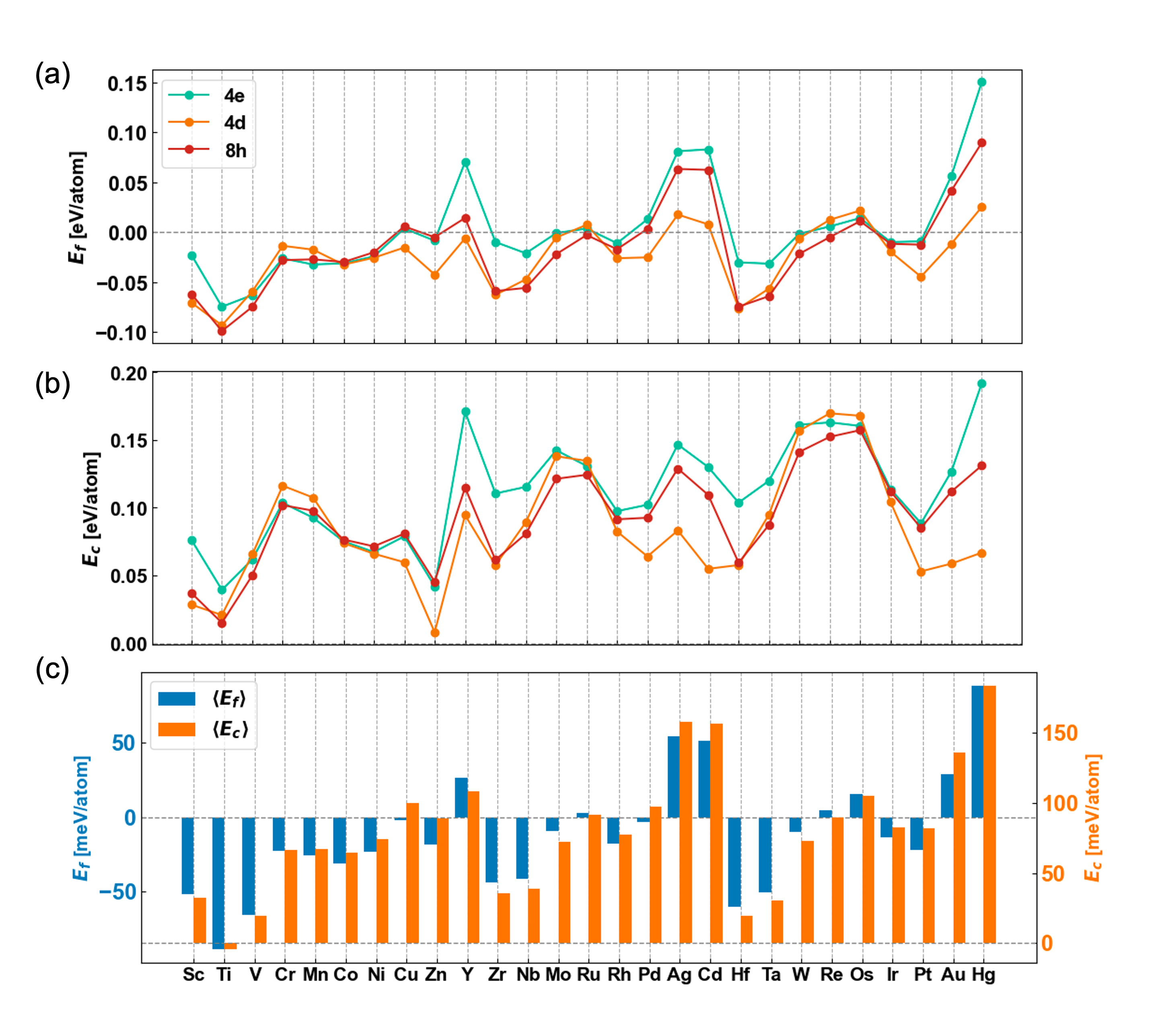}
  \caption{(a) Formation energy and (b) convex hull of $\mathrm{Fe_{15}XN_2}$ doped by different elements and iron sites. (c) Boltzmann-averaged formation energy and convex hull for each doping system.}
  \label{fig:fig2}  
\end{figure}

The thermodynamic stability of alloyed Fe$_{15}$XN$_2$ can be characterized by the formation energy $E_f$ and the distance to convex hull $E_c$, as shown in Fig.~\ref{fig:fig2}.
The formation energy is obtained by

\begin{align}
    E_f\bigl(\mathrm{Fe}_{15}\mathrm{XN}_{2}\bigr)
= & \frac{1}{18}E\bigl(\mathrm{Fe}_{15}\mathrm{XN}_{2}\bigr) \nonumber \\
& - \frac{15}{18}\,E(\mathrm{Fe})
- \frac{1}{18}\,E(\mathrm{X})
- \frac{2}{18}\,E(\mathrm{N}),
\end{align}
where $E\bigl(\mathrm{Fe}_{15}\mathrm{XN}_{2}\bigr)$ is the total energy of Fe$_{15}$XN$_2$, E(Fe), E(X), and E(N) are the energies of bcc Fe, the ground-state structure of the element X, and $\mathrm{N_2}$. The distance to convex hull $E_c$ is calculated by simply considering the pristine $\mathrm{Fe_{16}N_2}$, the Fe$_4$N, ground-state X and N as the competing phases given by
% \begin{align}E_c\bigl(\mathrm{Fe}_{15}\mathrm{XN}_{2}\bigr)
%     = & \frac{1}{18} E\bigl(\mathrm{Fe}_{15}\mathrm{XN}_{2}\bigr) \nonumber \\
%     & - \frac{15}{256} E(\mathrm{Fe_{16}N_2})
%     - \frac{1}{18} E(\mathrm{X}) \nonumber \\
%     & + (\frac{15}{256} - \frac{15}{18}) E(\mathrm{Fe}) + 
%     (\frac{15}{128} - \frac{2}{18}) E(\mathrm{N})
%     \label{eq:ch}
% \end{align}

\begin{align}E_c\bigl(\mathrm{Fe}_{15}\mathrm{XN}_{2}\bigr)
    = & \frac{1}{18} [E\bigl(\mathrm{Fe}_{15}\mathrm{XN}_{2}\bigr) \nonumber \\
    & - \frac{7}{8} E(\mathrm{Fe_{16}N_2})
    - \frac{1}{4} E(\mathrm{Fe_{4}N})
    \nonumber \\
    & - \frac{1}{8} E(\mathrm{Fe}) - E(X)]
    \label{eq:ch2}
\end{align}

Fe$_{15}$XN$_2$ is considered as thermodynamically stable when $E_c = 0$ and $E_f \le 0$, and metastable when $E_c$ is smaller than a chosen tolerance, e.g., $E_c \le$ 150 (meV/atom). %A lower $E_c$ from Eq.~\ref{eq:ch} suggests a higher likelihood of synthesizing the compound experimentally by doping $\mathrm{Fe_{16}N_2}$. 
% \textcolor{blue}{Please add the Boltzmann-averaged formation energy and convex hull}
Furthermore, to account for the site-preference, the Boltzmann average is considered to obtain the thermodynamic-averaged $E_f$ and $E_c$ for three possible doping sites.
Taking formation energy as an example, the averaged $E_f$ for doping element X is defined by
\begin{equation}
    \label{eq:mean}
    \langle E_f\rangle_X
    = \frac{\sum\limits_{i=4e,4d,8h}e^{-\Delta E/k_BT}\cdot E_f^i}{\sum\limits_{i=4e,4d,8h}e^{-\Delta E/k_BT}},
\end{equation}
where $\Delta E = E_{f}^i - E_f^{\mathrm{min}}$ is the energy difference between the formation energy of doping site i and the minimum doping formation energy. $k_B$ is Boltzmann constant and $T$ is the temperature chosen to be 300 K. 

The thermodynamic stabilities of Fe$_{15}$XN$_2$ show an interesting trend. 
According to Fig.~\ref{fig:fig2}(a), $E_f \le 0$ for most doping elements, except for Y, Ru, Ag, Cd, Au, Os, and Hg, $i.e.$, those with (almost) empty/full $d$-shell occupations. This gives rise to the positive Boltzmann-averaged $\braket{E_f}$ for these eight elements, as shown in Fig.~\ref{fig:fig2}, indicating Fe$_{15}$XN$_2$ (X = Y, Ru, Ag, Cd, Au, Os, and Hg) may decompose into the corresponding elements experimentally. Out of the 7 excluded, Ru, Re, and Os show slightly positive $E_f$ and can be stabilized under specific conditions. Thus, dopant elements are likely to be distributed over all three atomic sites at a reasonable synthesis temperature, necessitating a Boltzmann average for intrinsic magnetic properties.
It is noted that, although the 4d and 8h sites are generally favored, the energy difference among the 4e/4d/8h sites is under 40 meV/atom for 22 residual elements. Excluding cases with positive formation energy, 14 remain thermodynamically stable. In particular, Mn doping slightly leans towards 4e sites, aligning with Ref.~\cite{islam2020prediction}.
Finally, Fig.~\ref{fig:fig2}(b) and (c) indicate that $E_c$ and $\braket{E_c}$ are below 150 meV/atom for the 22 mentioned elements, with FeTiN showing a negative convex hull, making them viable for experimental synthesis. 
%\sout{As shown in Figure~\ref{fig:fig2} (c), most doped compounds fall within the metastable zone with negative $E_f$ and $E_c$ < 150 meV/atom, making them viable for experimental synthesis. We notice that in 4d and 5d groups, those elements lying in the "ends" (with empty or filled d orbitals) tend to have positive doping formation energy, e.g., Y, Ag, Cd, Au, and Hg. Site resolved stability in Figure~\ref{fig:fig2} (a) and (b) show that the 4d and 8h sites are generally preferred, but for Mn, the 4e site is the most favorable~\cite{islam2020prediction}.}

The trend observed in the thermodynamic stabilities can be understood based on the chemical bonding upon substitutions.
%\textcolor{red}{Please put down detailed COHP analysis here for at least two contrastive cases. The text you have below should be updated.}
An analysis of the integrated crystal Hamiltonian orbital population (ICOHP) in Table~S1 reveals that the Fe (4e)-N bonds exhibit the strongest bond strength with an -ICOHP of 2.55 eV, surpassing the Fe(8h)-N bonds, which have an -ICOHP of 1.82 eV. The 4d Fe, however, is not directly bonded with N. This accounts for the notable reduction in stability during 4e doping across these systems. Conversely, certain systems, such as $\mathrm{Fe_{15}ZrN_2}$, show a preference for 4d/8h site doping over 4e sites. The increased stability in the 8h site is primarily due to a shorter Fe (8h)-N bond length forming a 180\degree angle with the Zr-N bond (illustrated in Figure~S2 (c)), raising -COHP from 1.82 eV to 2.27 eV. In the case of 4e site doping, despite boosted Fe(4e)-N and Fe(8h)-N bond strengths, the significant weakening of the Zr(4e)-N bond by 2.25 eV (as seen by comparing Tables S1 and S3) is the key factor in reduced thermodynamic stability, making its formation energy higher than the other sites. Doping with Co and other 3d elements shows little site preference in stability due to their radii closely matching that of Fe. As confirmed by COHP analysis in Table~S4, the weakening of the X-N bond is minimal in these cases, hence its impact on neighboring Fe-N bonds is less significant.

\begin{figure}
  \centering
  \includegraphics[width=0.45\textwidth]{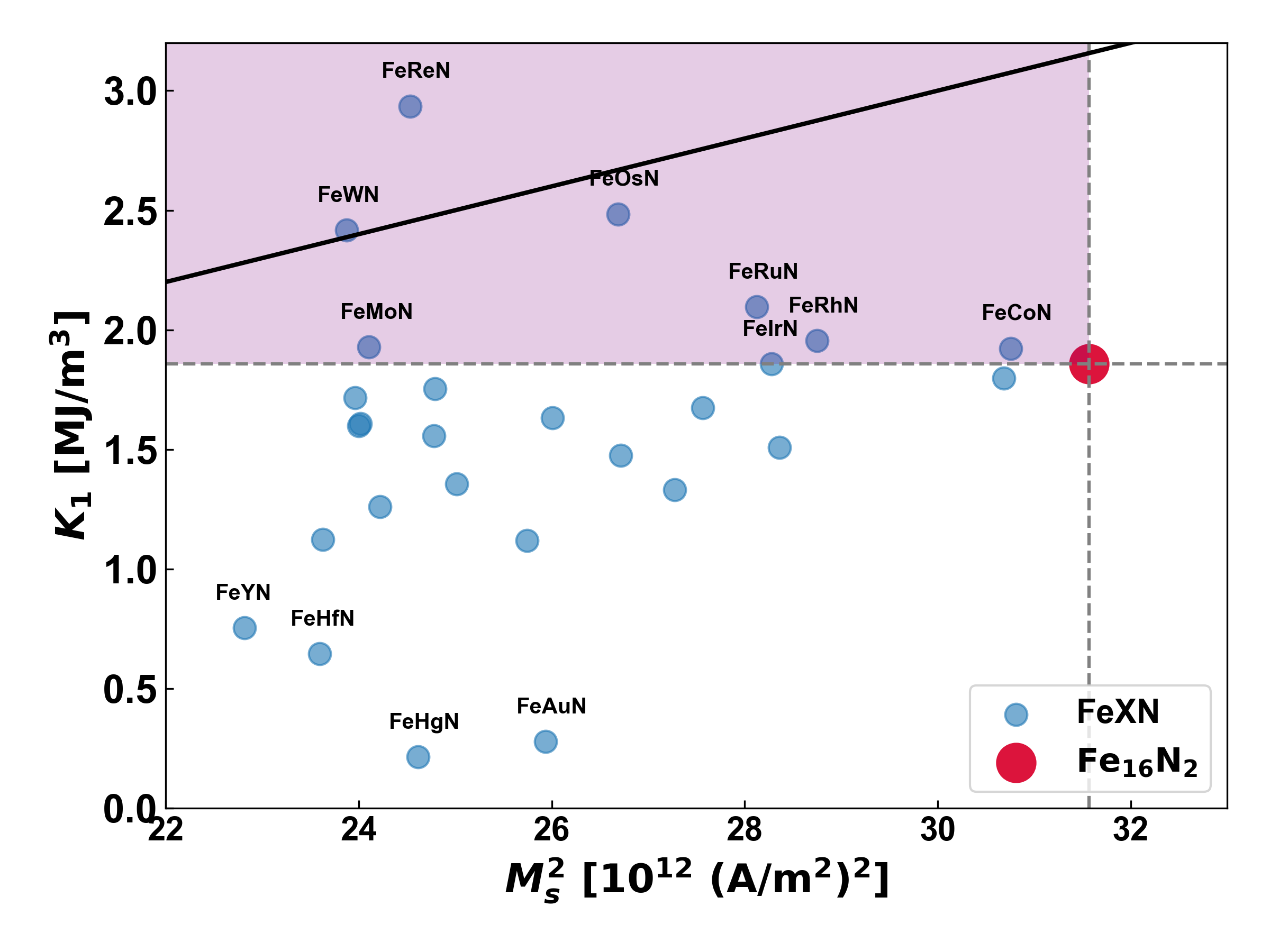}
  \caption{The MAE $K_1$ \textit{vs.} square of magnetization $M_s^2$ for doped $\mathrm{Fe_{15}XN_2}$ (X = d metals). The corresponding values for pristine $\mathrm{Fe_{16}N_2}$ are marked by the red dot. Region with $M_s^2 < M_s^2(\mathrm{Fe_{16}N_2})$ and $|K_1| > K_1 (\mathrm{Fe_{16}N_2})$ is marked by purple shade. The solid line corresponds to the magnetic hardness parameter $\kappa = \sqrt{\frac{K_1}{(\mu_0\mu_s)^s}}$ (here $\mu_0 = 0.1$ for SI unit system) for value $\kappa = 1$.
  %\textcolor{red}{SI unit as well}
  %\textcolor{blue}{If you set the x-range to be between 22 and 32, and the y-range to be between 0 and 32, we can zoom into the figure. Correspondingly, we can probably add labels for those 9 cases}
  %\textcolor{red}{Please put the site-resolved MAE-$M_S^2$ plot into the supplementary, b/c we are going to discuss the atom-resolved contributions.}
  }
  \label{fig:fig3}  
\end{figure}

To identify possible candidate permanent magnets, the MAE with respect to $M_S^2$ is shown in Figure~\ref{fig:fig3}. It is observed that all doped compounds show reduced $M_S$, as compared to that of pristine $\mathrm{Fe_{16}N_2}$. Interestingly, 8 out of 27 elements either maintain or enhance the MAE of $\mathrm{Fe_{16}N_2}$, as highlighted in the purple-shaded region in Figure~\ref{fig:fig3}. The corresponding intrinsic properties are listed in Table~\ref{tab:tab1}. Among them there are few 3d and 4d doping compounds that are $\mathrm{Fe_{15}CoN_2}$ with $K_1$ = 1.922 MJ/m$^3$, $\mathrm{Fe_{15}MoN_2}$ with $K_1$ = 1.928 MJ/m$^3$, and $\mathrm{Fe_{15}RhN_2}$ with $K_1$ = 2.097 MJ/m$^3$, indicating the possibility for light-element doping in enhacing MAE.
% Detailed information for these compounds is listed in Table~\ref{tab:fe_nitrides}, where stability can be evaluated by formation energy $\mathrm{E_{formation}}$ and Curie temperature $\mathrm{T_C}$. 
A dimensionless figure of merit~\cite{coey2014new} parameter $\kappa = \sqrt{K_1/(\mu_0\mu_s)^s}$ is defined to evaluate the performance of a permanent magnet, with $\kappa = 1$ threshold depicted as the solid line in Figure~\ref{fig:fig3}. According to $\kappa$, $\mathrm{Fe_{15}ReN_2}$, $\mathrm{Fe_{15}WN_2}$, and $\mathrm{Fe_{15}OsN_2}$ lie around the $\kappa = 1$ threshold, all by doping with 5d-elements with large atomic SOC strength.

% According to $\kappa$, six compounds stand out that are 4e-doped $\mathrm{Fe_{15}ReN_2}$ ($K_1 = 7.502\ \mathrm{MJ/m^3}$), 4d-doped $\mathrm{Fe_{15}TaN_2}$ ($K_1 = 2.874\ \mathrm{MJ/m^3}$), 4d-doped $\mathrm{Fe_{15}WN_2}$ ($K_1 = 2.993\ \mathrm{MJ/m^3}$), 4d-doped $\mathrm{Fe_{15}PtN_2}$ ($K_1 = 3.617\ \mathrm{MJ/m^3}$), 8h-doped $\mathrm{Fe_{15}OsN_2}$ ($K_1 = 3.509\ \mathrm{MJ/m^3}$), 8h-doped and $\mathrm{Fe_{15}IrN_2}$ ($K_1 = 4.946\ \mathrm{MJ/m^3}$), all by doping with 5d-elements with large atomic SOC strength.

%\sout{All 3d-element dopants and parts of the 4d- and 5d-dopants are found to exhibit negative formation energies, indicating the possibility for experimental synthesis.}
The Curie temperatures are slightly lowered upon doping, as shown in Table.~\ref{tab:tab1} for the eight promising candidates. This can be attributed to the reduction of exchange coupling $J_{ij}$ between iron atoms, as compared to that of pristine $\mathrm{Fe_{16}N_2}$ at 1273 K estimated by mean field approximation, which is higher than experimental $T_C$ around 810 K~\cite{sugita1991giant,takahashi1994magnetic}.
Nevertheless, $T_C$ is higher than 1000 K for all the eight cases. Furthermore, considering the slightly positive $E_f$ for Ru/Os/Ir doping and the price of elements, we identify Co, Mo, and W as the three most promising cases for further experimental validations.

\begin{table*}

    \centering
    \caption{Compounds with promising MAE. All properties listed are the Boltzmann average over sites calculated from Eq.~\ref{eq:mean}. 
    % \textcolor{blue}{Add Fe$_8$N as a reference. Please use the SI units for M$_s$ and K$_1$.}
    }
    \begin{tabular}{lrrrr}
        \toprule
        \textbf{Compound} & $\mathrm{\langle E_f\rangle}$ [meV/atom] & $\langle M_s \rangle \ [10^6\ \mathrm{A/m}]$ & $\langle T_C \rangle$ [K] & $\langle K_1 \rangle\ [\mathrm{MJ/m^3}]$ \\
        \midrule 
        FeCoN & -31.06 & 5.55 & 1260.43 & 1.922 \\
        FeMoN & -9.30  & 4.91 & 1103.70 & 1.928 \\
        FeRuN & 2.86   & 5.30 & 1096.90 & 2.097 \\
        FeRhN & -17.93 & 5.36 & 1139.23 & 1.955 \\
        FeWN  & -9.64  & 4.89 & 1113.77 & 2.417 \\
        FeReN & 4.53   & 4.95 & 1039.83 & 2.935 \\
        FeOsN & 15.75  & 5.17 & 1058.57 & 2.484 \\
        FeIrN & -13.58 & 5.32 & 1107.23 & 1.857 \\
        $\mathrm{Fe_{16}N_2}$ & -38.26 & 5.62 & 1273 & 1.857 \\
        \bottomrule
    \end{tabular}
    \label{tab:tab1}
\end{table*}

\begin{figure}[ht]
  \centering
  \includegraphics[width=0.49\textwidth]{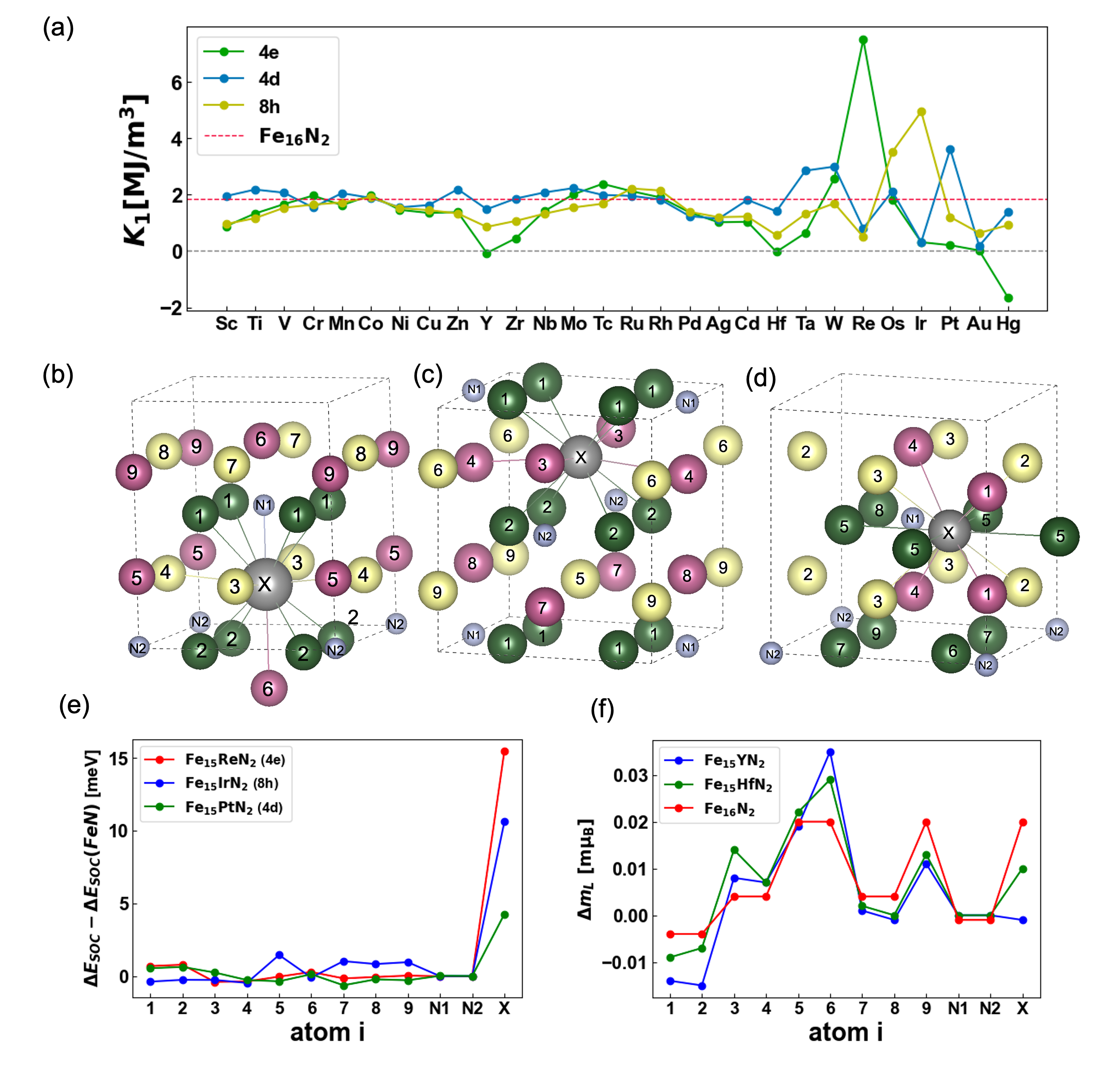}
  \caption{(a) Atom-resolved MAE for different doping site and doping atom; Local structure of (b) 4e-, (c) 4d-, and (d) 8h-doped $\mathrm{Fe_{15}XN2}$, X is the doping atom (iron atoms are labeled in the order of the distance to the doping atom X); (e) Difference between the SOC energy anisotropy ($\Delta E_{SOC}^{[100]-[001]}$) of $\mathrm{Fe_{15}XN_2}$ with enhanced out-of-plane MAE and the corresponding site in pristine $\mathrm{Fe_{16}N_2}$. Atoms are labeled corresponding to the local structures in (b)-(d). (f) The orbital moment anisotropy of atom X and the neighboring atoms for in-plane MAE doping systems (all 4e-site doping).}
  \label{fig:fig4}  
\end{figure}

To shed more light on the origin of enhanced/reduced MAE for the doped Fe$_{15}$XN$_2$, a detailed analysis of the site-resolved contributions is performed.
% \textcolor{blue}{Here I would suggest the following step-by-step analysis. 
% Step-1: compare the Boltzmann-averaged Figure 1 and the counterpart 4d/4e/8h-resolved Figure to discuss about the influence of different doping sites, using one case as an example where 4d/4e/8h-doping has a big difference in MAE.
% Step-2: Analyze on one case where the resulting MAE is (almost) negative
% Step-3: Analyze one of the promising candidate case as we identified above
% }
A direct comparison between the Boltzmann-averaged MAE of Figure~\ref{fig:fig3} and the site resolved values in Figure~\ref{fig:fig4} (a) reveals that the apparent trend across the dopants can mask pronounced site selectivity. 
% \textcolor{red}{When you are doing the analysis, please be quantitative, e.g., the values of $\Delta \mu_L$ and MAE should be explicitly specified, even though you are showing them in the plots.}
% Based on Bruno's theorem~\cite{bruno1989tight}, the contribution to MAE can be related to the orbital moment anisotropy $\Delta m_L = m_L^{\perp}-m_L^{\parallel}$ via $K_1 = \frac{\xi}{4}(m_L^{\perp}-m_L^{\parallel})$, with $\xi$ being the spin-orbit coupling constant.
We study some representative cases with significant changes of MAE by analyzing the SOC energies and orbital moments of the local structure at the doping site.
Ir and Pt, for example, display averaged MAEs that are only comparable to - or even below — those of pristine $\mathrm{Fe_{16}N_2}$ (1.86 and 1.68 MJ/m$^3$ respectively), yet their 8h-Ir and 4d-Pt configurations generate exceptionally large uniaxial anisotropies at 4.9 MJ/m$^3$ and 3.6 MJ/m$^3$ respectively as shown in Figure 4(a), signaling a strong preference for those sites; Re likewise shows a marked bias towards the 4e site of 7.5 MJ/m$^3$. Changes of SOC energy anisotropy $\Delta E_{SOC}$ ($E_{SOC}^{[100]} - E_{SOC}^{[001]}$) shown in Figure~\ref{fig:fig4}(e) trace these enhancements to a significant atomic SOC strength for the doping elements X (X = Re, Ir, and Pt), and the slight increase of $\Delta E_{SOC}$ in iron atoms at 8h site.
% Enormous negative $\Delta m_L$ are observed at the Re and Ir sites, which contradicts its large out-of-plane MAE, may indicate the failure of Bruno's theory where the spin-flip contribution to the MAE is neglected~\cite{van1998microscopic}. 
% The enormous negative $\Delta m_L$ at the Re and Ir sites, which contradicts its large out-of-plane MAE, may indicate the failure of Bruno's theory where the spin-flip contribution to the MAE is neglected~\cite{van1998microscopic}.

By contrast, 4e-substituted Hf and Y drive the MAE almost negative, both due to the fact that the 8h Fe atoms closest to the dopant acquire larger in-plane orbital moments, -14 m$\mu_B$ for Y and -9 m$\mu_B$ for Hf, respectively, while the dopant itself contributes little to $\Delta m_L$, as shown in Figure ~\ref{fig:fig4}(f). Among the most promising candidates—Co, Mo, and W—the Co impurity is notable for exhibiting virtually no site dependence in MAE, whereas Mo and W decouple thermodynamic stability and magnetic hardness: both metals are most stable at the 8h sites, yet their peak MAE occurs at 4d sites. Such decoupling is also observed in other systems comparing Figure~\ref{fig:fig2}(b) and Figure~\ref{fig:fig4}(a), highlighting an intrinsic trade-off that must be navigated in permanent-magnet designs~\cite{matsumoto2020sm,ke2016intrinsic}.

\section{Conclusion}

% \textcolor{blue}{The conclusions should be further extended --- as a rule of thumb, one summarizing per important paragraph in the result part, e.g., trend in the stability, no site-preference and thermodynamic average, COHP, 9 candidates and 3 most promising ones, origin of MAE based on site-resolved contributions. The most important numbers should be specified as well.}
In summary, we investigate the optimization of $\alpha''$-$\mathrm{Fe_{16}N_2}$ as permanent magnet via alloying with transition metals. Small $E_c$ (< 150 meV/atom) and negative $E_f$ are observed in 20 of the 27 substitution systems, making them viable for alloying. Site energy difference in less than 40 meV/atom is found in most systems, indicating no apparent site-biased stability. Variation of site stability can be understood by COHP analysis of chemical bonding.
By considering the averaged intrinsic magnetic properties ($M_s$, $K_1$, and $T_C$) at room temperature, eight systems are predicted to have enhanced MAE, within which three systems stand out as promising permanent magnets that are $\mathrm{Fe_{15}CoN_2}$, $\mathrm{Fe_{15}MoN_2}$, and $\mathrm{Fe_{15}WN_2}$, with corresponding $\langle K_1\rangle$ at 1.922 MJ/m$^3$, 1.928 MJ/m$^3$, and 2.417 MJ/m$^3$, respectively. Strong site preference of the properties are observed in some cases as a result of the interplay between the doping atom and the local environments. In particular, 4e Re, 8h Ir and 4d Pt are found to induce an enormous MAE due to the large atomic SOC strength of atom X (X=Re, Ir, Pt), while 4e Y and 4e Hf switch the direction of MAE from out-of-plane to in-plane by further enhancing negative $\Delta m_L$ of the neighboring 8h iron atoms. As exemplified by the case of $\mathrm{Fe_{15}MoN_2}$ and $\mathrm{Fe_{15}WN_2}$, a trade‐off between thermodynamic stability and magnetic anisotropy is worth investigating in permanent-magnet designs.

%%%%%%%%%%%% Supplementary Methods %%%%%%%%%%%%
%\footnotesize
%\section*{Methods}

%%%%%%%%%%%%% Acknowledgements %%%%%%%%%%%%%
%\footnotesize
\section*{Acknowledgements}

The authors gratefully acknowledge the computing time provided to them on the high-performance computer Lichtenberg at the NHR Centers NHR4CES at TU Darmstadt. This is funded by the Federal Ministry of Education and Research, and the state governments participating on the basis of the resolutions of the GWK for national high performance computing at universities (www.nhr-verein.de/unsere-partner). This work is funded by the Deutsche Forschungsgemeinschaft (DFG, German Research Foundation) - CRC 1487, "Iron, upgraded!" - with project number 443703006 and the DFG project with project number 471878653.

%%%%%%%%%%%%%%   Bibliography   %%%%%%%%%%%%%%
\normalsize
\bibliography{references}

%%%%%%%%%%%%  Supplementary Figures  %%%%%%%%%%%%
\clearpage
\section{Supplement Material}

% reset figure counter
\setcounter{figure}{0}
% redefine \thefigure to prefix “S”
\renewcommand{\thefigure}{S\arabic{figure}}
\setcounter{table}{0}
\renewcommand{\thetable}{S\arabic{table}}

\begin{figure}[H]
  \centering
  \includegraphics[width=0.48\textwidth]{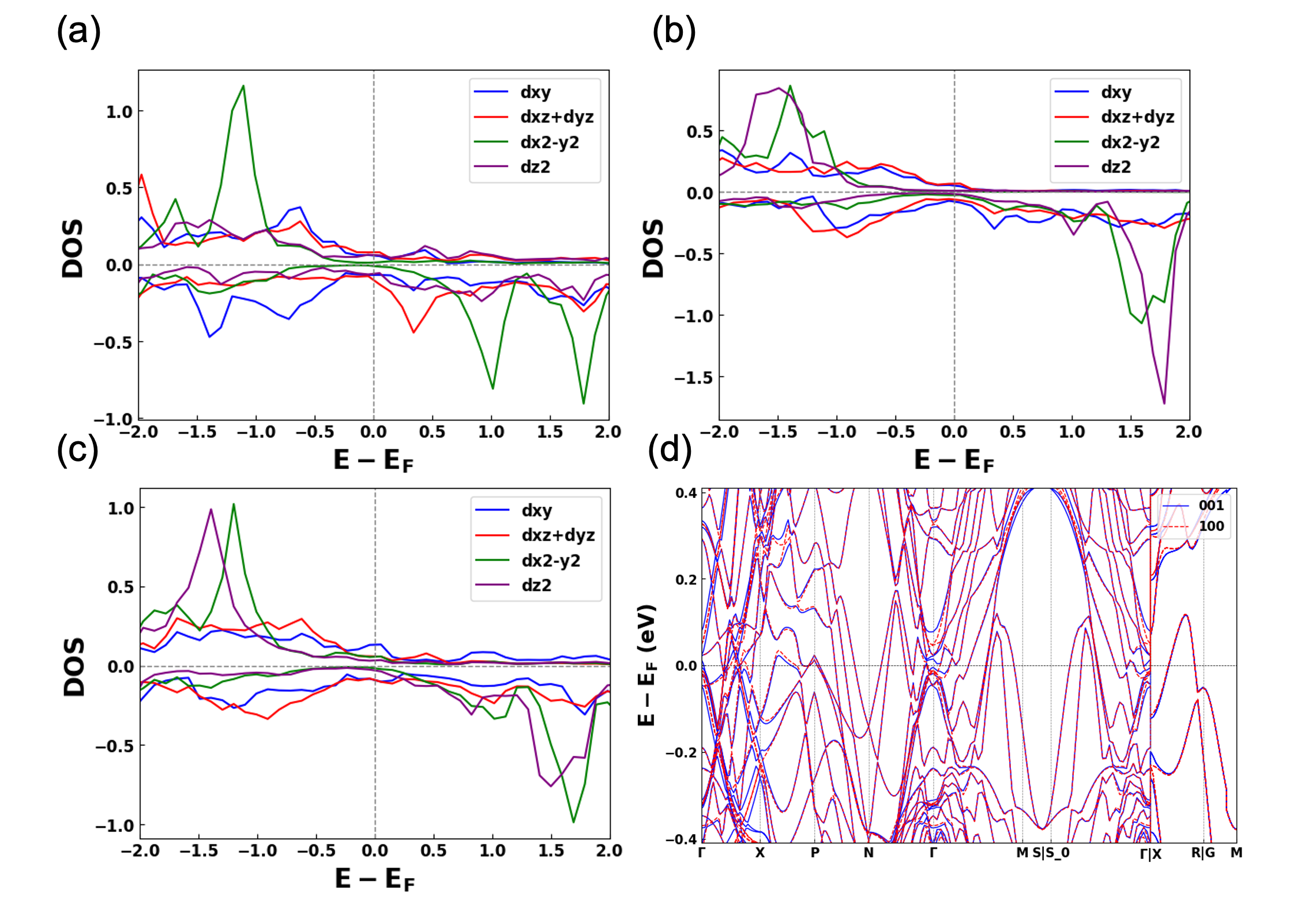}
  \caption{Density of states including SOC for iron atoms of (a) 4e, (b) 4d, and (8h) sites in $\mathrm{Fe_{16}N_2}$. In $D_{4h}$ symmetry, the d-orbitals are split into the following four irreducible representations: $A_{1g}\ (d_{z^2})$, $B_{1g}\ (d_{x^2-y^2})$, $B_{2g}\ (d_{xy})$, and $E_g\ (d_{xz}+d_{yz})$. (d) Band structure of Fe$_{16}$N$_2$ including SOC along quantization directions [001] (blue solid line) and [100] (red dashed line).}
  \label{fig:dos}  
\end{figure}

\begin{figure}[H]
  \centering
  \includegraphics[width=0.48\textwidth]{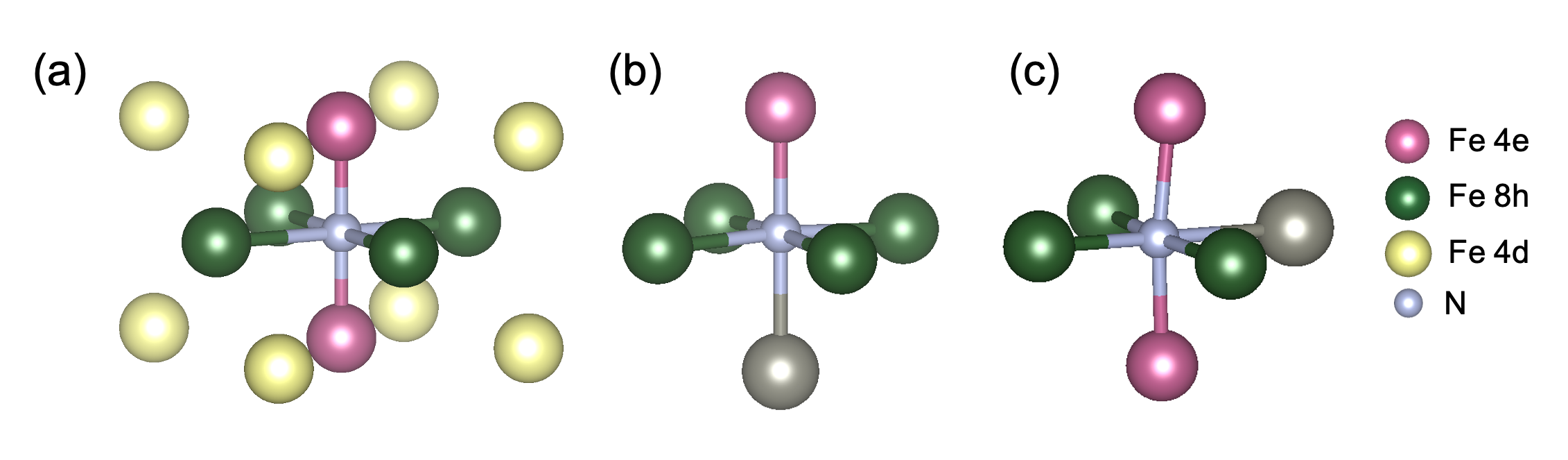}
  \caption{Local bonding environments (Fe$_{15}$ZrN$_2$ as an example) of (a) pristine $\mathrm{Fe_{16}N_2}$, (b) 4e site doping and (c) 8h site doping.}
  \label{fig:bonds}  
\end{figure}

\begin{table}[H]
    \centering
    \label{tab:tabs1}
    \caption{Bond length and strength given by -COHP values for bonds in pristine $\mathrm{Fe_{16}N_2}$}
    \begin{tabular}{lrr}
        \toprule
        \textbf{Bond} & distance [\AA] & -ICOHP [eV] \\
        \midrule
        Fe(4e)-N & 1.83 & 2.55 \\
        Fe(8h)-N & 1.95  & 1.82  \\
        Fe(4d)-Fe(8h) & 2.54  & 1.07 \\
        Fe(4d)-Fe(4e) & 2.85  & 0.43 \\
        \bottomrule
    \end{tabular}
\end{table}

\begin{table}[H]
    \centering
    \label{tab:tabs2}
    \caption{Bond length and strength given by -COHP values for bonds in 8h-doped $\mathrm{Fe_{15}ZrN_2}$, corresponding to Figure~S2 (c).}
    \begin{tabular}{lrr}
        \toprule
        \textbf{Bond} & distance [\AA] & -ICOHP [eV] \\
        \midrule
        Fe(4e)-N & 1.87 & 2.47 \\
        Fe(8h)1-N & 1.99  & 1.92  \\
        Fe(8h)2-N & 1.89  & 2.27 \\
        Zr(8h)-N & 2.17  & 0.24 \\
        \bottomrule
    \end{tabular}
\end{table}

\begin{table}[H]
    \centering
    \label{tab:tabs3}
    \caption{Bond length and strength given by -COHP values for bonds in 4e-doped $\mathrm{Fe_{15}ZrN_2}$}
    \begin{tabular}{lrr}
        \toprule
        \textbf{Bond} & distance [\AA] & -ICOHP [eV] \\
        \midrule
        Fe(4e)-N & 1.79 & 2.70 \\
        Fe(8h)-N & 1.99  & 1.96  \\
        Zr(4e)-N & 2.05  & 0.30 \\
        \bottomrule
    \end{tabular}
\end{table}

\begin{table}[H]
    \centering
    \label{tab:tabs4}
    \caption{Bond length and strength given by -COHP values for bonds in 4e-doped $\mathrm{Fe_{15}CoN_2}$}
    \begin{tabular}{lrr}
        \toprule
        \textbf{Bond} & distance [\AA] & -ICOHP [eV] \\
        \midrule
        Fe(4e)-N & 1.84 & 2.34 \\
        Fe(8h)-N & 1.94  & 2.037  \\
        Co(4e)-N & 1.83  & 1.10 \\
        \bottomrule
    \end{tabular}
\end{table}

\begin{figure}[H]
  \centering
  \includegraphics[width=0.48\textwidth]{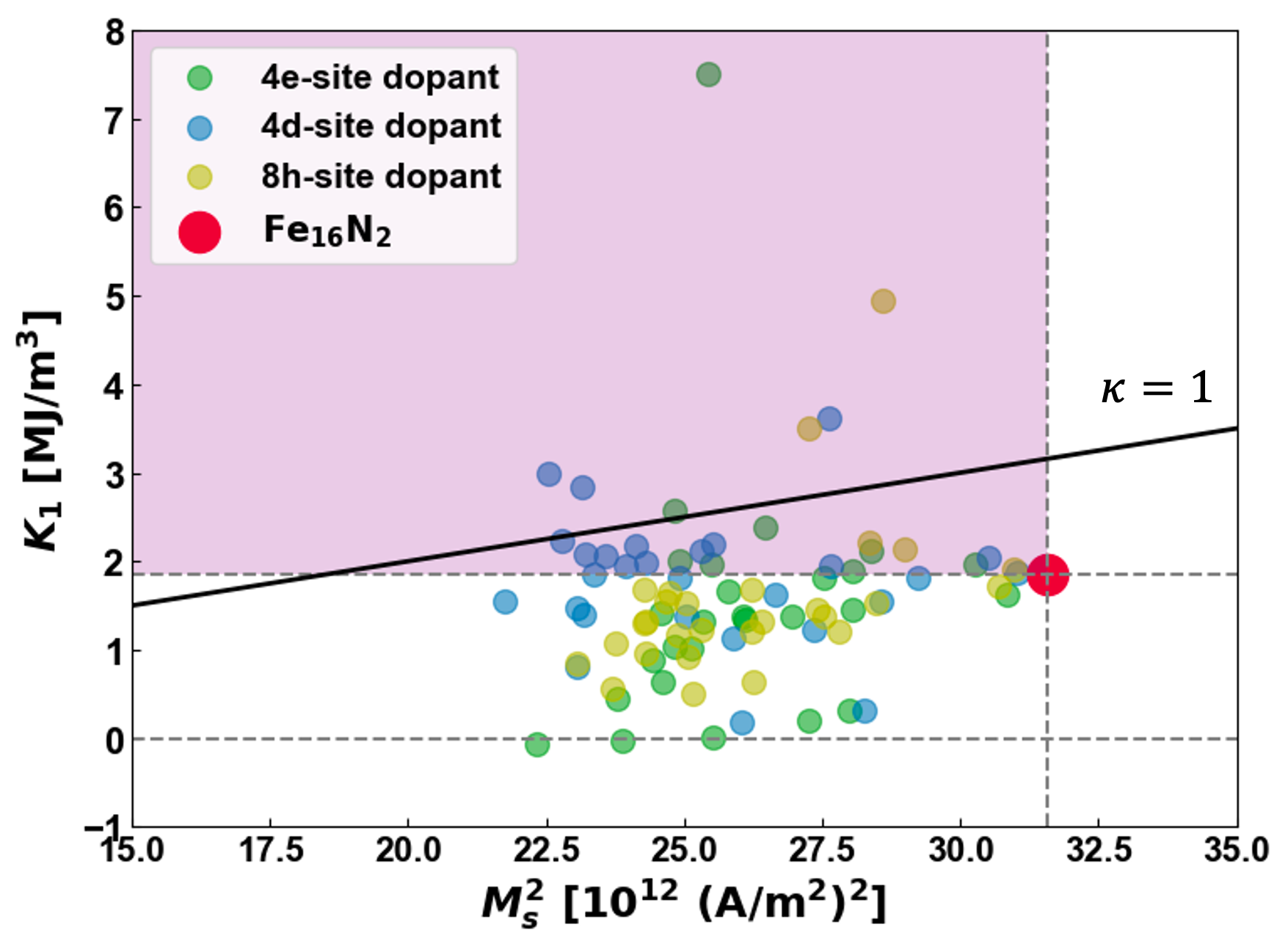}
  \caption{Site-decomposed $K_1$ \textit{vs.} square of magnetization $M_s^2$ for doped $\mathrm{Fe_{15}XN_2}$ (X=all d metals). The dashed line with magnetic hardness parameter $\kappa = 1$ is obtained in the same way as in Figure~3.}
  \label{fig:figs1}  
\end{figure}

\begin{figure}[H]
  \centering
  \includegraphics[width=0.48\textwidth]{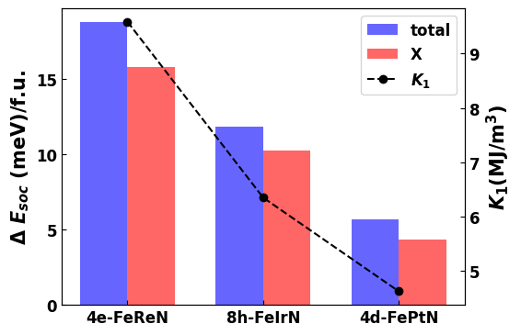}
  \caption{MAE (dashed line) vs SOC energy anisotropy $\Delta E_{SOC}$, i.e, $E_{SOC}^{[100]} - E_{SOC}^{[001]}$ (meV per formula unit) of both the structure (blue) and the doping elements (red), Re, Ir, and Pt.}
  \label{fig:figs2}  
\end{figure}

\begin{table}[H]
\centering
\caption{Calculated properties of transition‐metal films}
\label{tab:tm_props}
\begin{tabular}{lrrrr}
\toprule
Element & $\langle E_f\rangle$      & $\langle M_s\rangle$         & $\langle K_1\rangle$           & $\langle T_c\rangle$  \\
        & [meV/atom] & [$10^6$ A/m]  & [MJ/m$^3$]      & [K]    \\
\midrule
Sc &  -51.70 &  4.92 &  1.26 & 1162.37 \\
Ti &  -88.87 &  4.98 &  1.56 & 1200.13 \\
V  &  -65.57 &  4.98 &  1.75 & 1214.00 \\
Cr &  -22.51 &  4.89 &  1.72 & 1151.87 \\
Mn &  -25.55 &  5.54 &  1.80 & 1134.47 \\
Co &  -31.06 &  5.55 &  1.92 & 1260.43 \\
Ni &  -23.21 &  5.33 &  1.51 & 1144.80 \\
Cu &   -1.86 &  5.17 &  1.48 & 1133.03 \\
Zn &  -18.53 &  5.10 &  1.63 & 1159.60 \\
Y  &   26.25 &  4.78 &  0.75 & 1109.73 \\
Zr &  -43.76 &  4.86 &  1.12 & 1163.97 \\
Nb &  -41.25 &  4.90 &  1.61 & 1165.27 \\
Mo &   -09.30 &  4.91 &  1.93 & 1103.70 \\
Tc &   -00.04 &  5.06 &  2.01 & 1042.90 \\
Ru &    02.86 &  5.30 &  2.10 & 1096.90 \\
Rh &  -17.93 &  5.36 &  1.95 & 1139.23 \\
Pd &   -02.87 &  5.22 &  1.33 & 1109.23 \\
Ag &   54.15 &  5.07 &  1.12 & 1130.13 \\
Cd &   51.20 &  5.00 &  1.36 & 1161.10 \\
Hf &  -60.27 &  4.86 &  0.65 &  --     \\
Ta &  -50.56 &  4.90 &  1.60 & 1168.03 \\
W  &   -09.64 &  4.89 &  2.42 & 1113.77 \\
Re &    04.53 &  4.95 &  2.94 & 1039.83 \\
Os &   15.75 &  5.17 &  2.48 & 1058.57 \\
Ir &  -13.58 &  5.32 &  1.86 & 1107.23 \\
Pt &  -21.96 &  5.25 &  1.68 & 1110.17 \\
Au &   28.94 &  5.09 &  0.28 & 1118.57 \\
Hg &   88.51 &  4.96 &  0.21 & 1041.93 \\
\bottomrule
\end{tabular}
\end{table}

%%%%%%%%%%%%%%%%   End   %%%%%%%%%%%%%%%%
%\end{multicols}  % Method B for two-column formatting (doesn't play well with line numbers), comment out if using method A
\end{document}